# Super-resolution-based Change Detection Network with Stacked Attention Module for Images with Different Resolutions

Mengxi Liu[1]., Qian Shi[1*]., Andrea Marinoni[2,3]., Da He[1]., Xiaoping Liu[1]., Liangpei Zhang[4].

[1]. School of Geography and Planning, and Guangdong Provincial Key Laboratory for Urbanization and Geo-simulation, Sun Yat-sen University, Guangzhou, China

[2]. Dept. of Physics and Technology, UiT the Arctic University of Norway, Tromsø, Norway

[3]. Dept. of Engineering, University of Cambridge, Cambridge, UK

[4]. State Key Laboratory of Information Engineering in Surveying, Mapping and Remote Sensing, Wuhan University

*. Corresponding author (shixi5@mail.sysu.edu.cn)

**Abstract:** Change detection, which aims to distinguish surface changes based on bi-temporal images, plays a vital role in ecological protection and urban planning.  Since high resolution (HR) images can not be typically acquired continuously over time, bi-temporal images with different resolutions are often adopted for change detection in practical applications. Traditional subpixel-based methods for change detection using images with different resolutions may lead to substantial error accumulation when HR images are employed; this is because of intraclass heterogeneity and interclass similarity. Therefore, it is necessary to develop a novel method for change detection using images with different resolutions, that is more suitable for HR images. To this end, we propose a super-resolution-based change detection network (SRCDNet) with a stacked attention module. The SRCDNet employs a super resolution (SR) module containing a generator and a discriminator to directly learn SR images through adversarial learning and overcome the resolution difference between bi-temporal images. To enhance the useful information in multi-scale features, a stacked attention module consisting of five convolutional block attention modules (CBAMs) is integrated to the feature extractor. The final change map is obtained through a metric learning–based change decision module, wherein a distance map between bi-temporal features is calculated. The experimental results demonstrate the superiority of the proposed method, which not only outperforms all baselines -with the highest F1 scores of 87.40% on the building change detection dataset and 92.94% on the change detection dataset -but also obtains the best accuracies on experiments performed with images having a 4× and 8× resolution difference. The source code of SRCDNet will be available at https://github.com/liumency/SRCDNet.

# 1. Introduction

Change detection (CD) aims to identify surface changes in bi-temporal images of the same area and provides quantitative data for various important applications [1], such as land and resource investigation, ecological monitoring and protection, and urban planning [2-4]. Because of their ability to provide rich information on wide areas of Earth surface with high temporal efficiency, remote sensing images have been widely used in these applications for a long time [5].

In the past few decades, multi-spectral satellite images have been extensively employed in remote sensing applications [6, 7], including change detection [8]. Therefore, traditional change detection methods, such as change vector analysis (CVA) [9] and principle component analysis (PCA) [10], mainly exploit the spectral information in bi-temporal images. However, as the spectral and spatial resolutions of an image are mutually restricted, multi-spectral satellite images usually have a low spatial resolution, making it difficult to achieve precise change detection. Recently, with the rapid development of remote sensing technology, high-resolution (HR) images with rich spatial information have become the main data source for change detection, especially for fine-grained scenarios such as urban renewal [11]. In order to fully exploit the opportunities provided by HR images, advanced techniques for remote sensing data analysis have been explored and proposed in technical literature. Deep learning–based methods that include a powerful feature learning structure, namely, convolutional neural networks (CNNs), to extract spatial and semantic features from HR images hierarchically [12] have provided a remarkable solution for high-resolution image change detection. Examples of these methods include classification-based methods [13-15] and metric learning-based methods [16, 17].

Although HR images might help in improving the characterization of the phenomena occurring on Earth surface, it is also true that several meteorological and technical effects – such as small observation range, low temporal resolution, and the effect of clouds and fog – might limit the actual capacity and impact of the investigation of HR images. Hence, the use of traditional change detection methods relying on the analysis of bitemporal images showing same resolution properties might not be adequate to tackle the main issues of remote sensing-based change detection, especially on a large scale scenarios. For example, let us suppose that we have obtained an HR image of a certain area at time *T1*, but only low-resolution (LR) images corresponding to time *T2* are available; then, we need to detect the changes that occurred between *T1* and *T2* using the bi-temporal images with different resolutions. Therefore, to realize large-scale and rapid change detection, it is often necessary to use bi-temporal images with different resolutions for real-life applications [18-20].

To tackle these issues, the most intuitive method is to simply downsample the HR image to the resolution of the LR image [21], or to interpolate the LR image to the resolution of the HR image to obtain bi-temporal images with the same resolution [22], and then employ common change detection methods to detect changes. However, the downsampling step at the core of the first approach induces a lack of detailed spatial information of the outcomes, which leads to a strong degradation of the precision of the obtained results. On the other hand, the second approach does not take into account semantic information in common interpolation operations (such as linear, bilinear, and bicubic interpolation) applied to remote sensing images, which leads to a scarce capacity to achieve detailed information on the region of interest.

In addition to the simplest interpolation methods mentioned above, other methods have been proposed to solve the problem of change detection with remote sensing images of different resolutions: in this context, the subpixel-based method is the most prevalent. Considering the excellent ability of sub-pixel mapping (SPM) to obtain fine-resolution land-cover maps from coarse-resolution images [23-25], Ling et al. [26] first introduced SPM to change detection using images with different resolutions (henceforth referred to as "different-resolution change detection" for brevity) using the spatial dependence principle and a novel land-cover change (LCC) rule to obtain the spatial pattern of LCC maps at the subpixel scale. Later, Wang et al. [27] proposed a Hopfield neural network with SPM to overcome the resolution difference between Landsat and MODIS images for subpixel-resolution land-cover changes. Li et al. [28] applied an iterative super-resolution change detection method for Landsat-MODIS change detection, which combines end-member estimation, spectral unmixing, and SPM. Wu et al. [29] proposed a back propagation neural network to obtain subpixel land-cover change maps from the soft-classification results of LR images.

These SPM-based methods obtain fine-resolution change maps by establishing a mapping between a former fine-resolution land-cover map and a coarse-resolution image and have been proven to be effective in dealing with large-scale differences for remote sensing image change detection, especially on Landsat and MODIS images. However, in these cases, the accuracy of fine-resolution change maps is largely limited by the accuracy of the former fine-resolution land-cover map, leading to redundant error accumulation. Such a problem would be much more severe for HR images due to the intraclass heterogeneity and interclass similarity in HR images. Therefore, there is an urgent need to develop more precise change detection methods for HR images with different spatial resolutions.

In this paper, we propose an end-to-end super-resolution-based change detection network (SRCDNet) for remote-sensing images with different resolutions. To deal with the unmatched spatial resolution of bi-temporal images, the SRCDNet employs a super-resolution module to learn an SR image directly from the LR images to recover more semantic information and avoid redundant errors. The super-resolution (SR) image is then input into a feature extractor together with the HR images corresponding to other timestamps. To fully extract the multi-level information in HR images, so as to facilitate the subsequent prediction, a stacked attention module consisting of five convolutional block attention modules (CBAMs) is also added to the feature extractor. Then, in order to learn precise change maps from the the multi-scale features of the bi-temporal images, the distance map between the features is calculated and compared with the ground truth, where a contrastive loss, a common loss in metric learning, is adopted to help increase the distance of the changed area and decrease that of the unchanged area. Finally, the change map could be obtained from the distance map through simple thresholding. Thus, the main contributions of this study can be summarized as follows:

(1) We provide an end-to-end super-resolution-based network for high-resolution image change detection; the proposed scheme learns the SR image through the mapping between the LR image and the initial HR image to avoid the error accumulation encountered in traditional subpixel-based methods.

(2) We integrate a stacked attention module consisting of five CBAM blocks into the feature extractor of the network to enhance valid information in the hierarchical features for more distinguishable feature pairs, which can greatly help the subsequent change decision through metric learning.

(3) We conducted change detection experiments on images with different scales, including initial HR images, images with four times (4×) resolution difference and those of eight times (8×) resolution difference, to explore the influence of images with varied scales on change detection results. Compared with other baseline methods, our method not only achieved the best performance on the initial HR images on both the building change detection dataset (BCDD) (Ji et al. 2018) and change detection dataset (CDD) [31], with F1 scores of 87.40% and 92.94%, respectively, but also achieved the highest accuracies in the two LR experiments.

## 2. Related Work

### 2.1. Deep learning–based change detection

After the proposal of a fully convolutional network (FCN) [32] provided a more intuitive method for dense prediction, many methods based on FCNs and their variants, especially U-Net [33], have been proposed for pixel-wise change detection. Daudt et al. [14] explored three different methods of image input, including early fusion, Siamese difference, and Siamese concatenation, based on U-Net for bi-temporal change detection. To fully utilize both global and local information in bi-temporal images, Peng et al. [34] proposed U-Net++ with a multi-scale feature fusion strategy to generate final change maps. The encoder–decoder structure of U-Net is commonly used in semantic segmentation tasks, where the encoder is used to extract multi-level semantic features of bi-temporal images, and the decoder is used to recover spatial information from the hierarchical features and generate change detection maps by classification.

In recent years, some studies have introduced metric learning into change detection to replace the decoder's upsampling process, which directly obtains change maps by calculating the distance between the features of the bi-temporal images. During the training process, the distance between "unchanged" features is minimized, while that between "changed" features is maximized; the loss function plays an important role in this process. For example, Zhang et al. [35] used an improved triplet loss to learn the semantic relation between multi-scale information in paired features. Wang et al. (Zhang et al. 2018) employed contrastive loss to detect changes based on features extracted using a Siamese convolutional network. To mitigate the effects of class imbalance in change detection, Chen et al. [16] employed batched contrastive loss to train the proposed spatial-temporal attention-based network (STANet).

### 2.2. Change detection strategies

While HR images critically lead to false alarms or missed alarms due to the influence of intraclass heterogeneity and interclass similarity, many attempts have been made to generate more discriminative features, including recurrent neural networks (RNNs) [37] and attention mechanisms. Papadomanolaki et al. [13] integrated long short-term memory blocks (LSTMs) (Papadomanolaki et al. 2019) into a fully convolutional network to detect urban changes in Sentinel-2 images, and proved that RNNs are effective for capturing spectral or temporal relationships between images. In addition, Song et al. [39] combined a convolutional LSTM with a 3D fully convolutional network to capture joint spectral–spatial–temporal features in hyperspectral images. Although RNNs can work well on multi-spectral and hyperspectral inputs, they are still limited because of infrequent spectral information in HR images and deficient time information in bi-temporal images.

Owing to their ability to enhance useful information in extracted features, attention mechanisms have been adopted to make better use of the abundant spatial information in HR images. Chen et al. (Chen and Shi 2020) integrated a self-attention module in the feature extractor of a CD network to strengthen the spatial-temporal relationships between bi-temporal features. Chen et al. [17] used dual CBAMs [40] for each bi-temporal feature to emphasize change information in images, facilitating subsequent metric learning–based change decisions. However, because of memory limitations, in the existing methods in which spatial information is effectively exploited for change detection, the attention mechanism is usually only applied to high-level semantic features, whereas the enhancement of shallow features is ignored.

### *2.3.  Super resolution*

To better restore the detailed information of an image when reconstructing HR images from LR images, a series of effective SR algorithms have been successively proposed by researchers. Dong et al. [41] first introduced CNNs into SR applications as SRCNN networks. Until now, based on deep learning technology, SR image reconstruction has achieved remarkable success. By combining an SRCNN and a VGG backbone, Kim et al. [42] designed a deep neural network called VDSR with 20 layers. Kim et al. [43] also proposed a deep recursive convolutional network (DRCN) for SR. In view of the excellent performance of generative adversarial networks (GANs) [44] in various other fields, Ledig et al. [45] applied a GAN to SR and proposed SRGAN, which achieved state-of-the-art performance at the time.

## 3. Methodology

In this section, a brief overview of the proposed method is provided, after which a detailed description of each part of the model is provided, and the optimization process of the model is presented.

### *3.1.  Overview*

The SRCDNet, as shown in Fig. 1, consists of two parts: an SR module to reconstruct the LR images of the bi-temporal images to HR images, and a change detection module to generate change maps. Based on a GAN, the SR module consists of a generator and a discriminator, whereas the change detection part of the SRCDNet contains a feature extractor to extract multi-scale features from the bi-temporal inputs for subsequent change decisions based on deep metric learning.

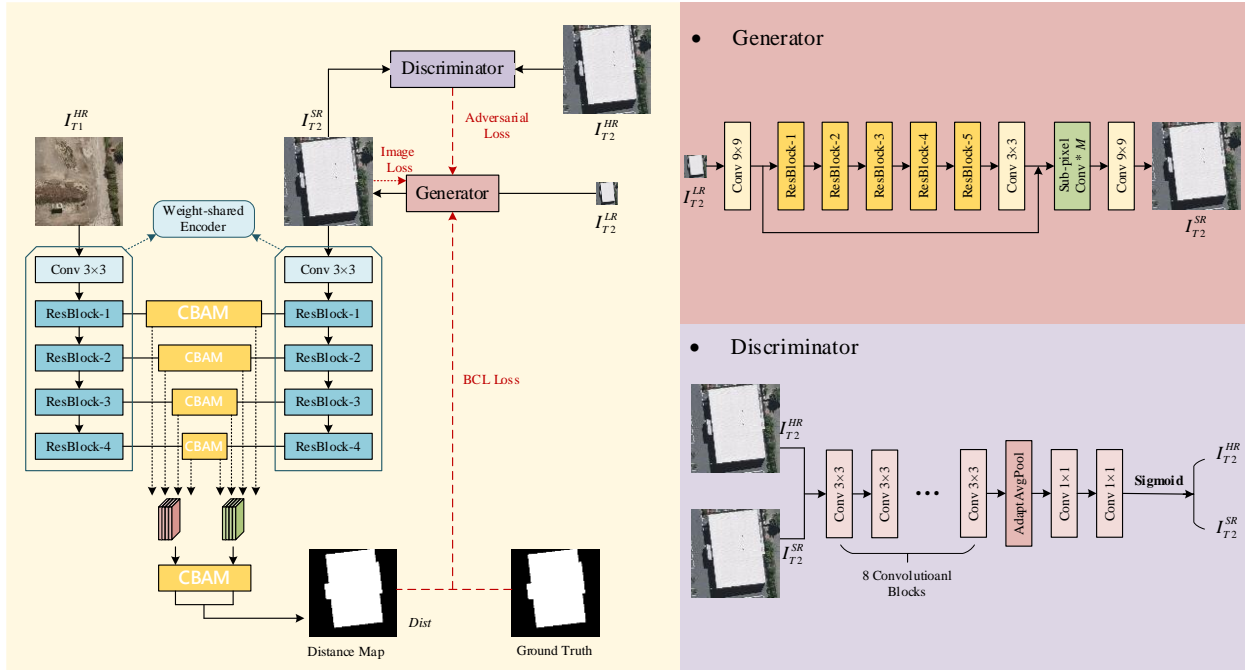

**Fig. 1. Overview of the proposed SRCDNet.**

Let us assume that we need to detect changes between HR images obtained at *T1* and LR images obtained at *T2*, $I_{T1}^{HR}$ and $I_{T2}^{LR}$, respectively. Moreover, let the resolution difference between a HR image and a LR image be equal to a factor *N* (*N* = 4, 8). Thus, for a set of training sets containing HR images $I = \{(I_{T1}^1, I_{T2}^1), \ldots, (I_{T1}^n, I_{T2}^n)\}, n \in N$ and the corresponding ground truth $Y = \{Y^1, \ldots, Y^n\}, n \in N$, the flowchart of SRCDNet can be summarized as follows:

(1) First, the HR image at *T2* $I_{T2}^{HR}$ would be downsampled *N* times to an LR image to simulate the LR image at *T2* $I_{T2}^{LR}$, which would be input into the SR module.

(2) In the SR module, the generator *G* produces an SR image $I_{T2}^{SR} = G(I_{T2}^{LR})$ with the same size as $I_{T2}^{HR}$ according to $I_{T2}^{LR}$; after that, the discriminator *D* is responsible for learning to discriminate $I_{T2}^{SR}$ from $I_{T2}^{HR}$ through the loss $Loss_D$, which is constituted by the output of the discriminator, $D(I_{T2}^{HR})$ and $D(I_{T2}^{SR})$.

(3) Then, the HR image obtained at *T1* $I_{T1}^{HR}$ and the SR image obtained at *T2* $I_{T2}^{SR}$ are both fed to the weight-shared feature extractor to extract hierarchical features. Four convolutional block attention modules (CBAMs), a lightweight attention mechanism that can enhance the features both channel-wisely and spatial-wisely, are applied on four intermediate features, before being stacked into one to obtain features with multi-scale information of both timestamps, $f_{T1}^{HR}$ and $f_{T2}^{SR}$, on which the fifth CBAM blocks would be applied.

(4) Thereafter, a distance map *Dist* is calculated based on the bi-temporal features $f_{T1}$ and $f_{T2}$ to measure the distance between $I_{T1}^{HR}$ and $I_{T2}^{SR}$, which would be compared with the ground truth and obtain a contrastive loss, $Loss_{CD}$. The metric $Loss_{CD}$ would be then optimized to push away the distance between the changed area on the ground truth and pull in the distance between the unchanged area.

(5) Finally, the generator $G$ is optimized according to the difference between the SR image $I_{T2}^{SR}$ and the HR image $I_{T2}^{HR}$, the discriminator's result, and the change detection performance $Loss_{CD}$, to generate SR images with rich semantic information.

It is worth noting that when there is no spatial resolution difference between bi-temporal images the SR module in SRCDNet can be easily removed, and the proposed network is then reduced to a simple change detection network (CDNet); this detail significantly improves the generality of the model.

### 3.2. Super resolution module

The structure of the SR module is inspired by the the SRGAN scheme [45], where a generator is responsible for generating the SR image from the LR image, whereas the discriminator distinguishes the SR image from the initial HR image until they are indistinguishable from each other and the generator is able to output SR images that are sufficient for fine-grained change detection.

The generator first employs a 9 × 9 convolutional layer to capture the shallow features of the input LR image, and then five residual blocks to extract high-level features. Each residual block is composed of two 3 × 3 convolutional layers followed by a batch normalization (BN) layer [46], of which the first one is a PReLU layer that serves as the activation layer. The deep features further extracted by residual blocks are fused with the shallow features obtained in the first convolutional layer to obtain features with rich spatial and semantic information. Consequently, $M$ sub-pixel convolution layers are used to increase the feature size to equal that of the HR image, where $M = log_2 N$. Finally, the generator produces an SR image by means of a fully convolutional layer.

Inheriting the structure of VGG-19 [47], the discriminator contains eight convolutional layers, wherein BN layers and leaky ReLU functions are used. Two fully connected layers and a sigmoid function are used to output the distribution of the input image, which is a binary classification task. Because the objective of the discriminator is to distinguish SR images from HR images, the generator can be prompted to generate SR samples that are more similar to the original HR image through adversarial training.

### 3.3. Change detection module

While the SR module aims to produce SR images similar to HR images, the task of the change detection module is to generate precise change maps based on the SR image and the HR image at another timestamp. The change detection module in the SRCDNet employs metric learning to obtain change maps based on features extracted from the feature extractor.

We use a pretrained ResNet-18 [40] as the feature extractor after removing the last fully connected layers, which is extended to a Siamese structure to receive bi-temporal inputs. A 7 × 7 convolutional layer with a stride of 1 is used to extract the shallow features with rich spatial information, followed by a BN layer and the ReLU function, after which a max-pooling layer with a stride of 2 is employed. Then, four residual blocks are employed to fully exploit the information in the images. The size of the output features of each residual block is 1/2, 1/4, 1/8, and 1/8 of the input image, and the channel of them are 64, 128, 256, and 512, respectively.

To fully capture the effective information in multi-scale features, we integrated a stacked attention module with five CBAMs in the feature extractor. More specifically, four CBAM blocks are applied to the output features of each residual block to emphasize useful information; these features are uniformly adjusted to 1/2 the size of the original image, and then fused into ones with multi-scale information. Thereafter, the fifth CBAM block is applied to the feature pairs to make more distinguishable for subsequent detection.

Each CBAM block contains two parts: a channel attention module to capture channel-wise relationship and a spatial attention module to explore spatial-wise contextual information. Given a feature $F$ with size $C \times H \times W$, an average pooling layer and a max pooling layer are first applied on the input feature respectively to obtain two vectors with size $C \times 1 \times 1$ in the channel attention module, then a weight sharing multi-layer perception (MLP) module with two 1×1 convolutional layers is used to learn and give weights to each channel. Finally, the two factors are summed up into one and a sigmoid function $\sigma$ is applied to get the channel attention map factor, which can be expressed as:

$$M_c(F) = \sigma(MLP(Avg(F)) + MLP(Max(F))) \tag{1}$$

The chanel-refined feature $F'$ is the result of multiplication of $M_c(F)$ and $F$, which can be denoted as:

$$F' = M_c(F) \otimes F \tag{2}$$

Thereafter, the spatial attention module would be applied on feature $F'$ with size of $C \times H \times W$, which is the same with $F$. Here, an average pooling layer and a max pooling layer are utilized to squeezed the $F'$ into two matrixes of size $1 \times H \times W$, which are then stacked into one and input into a 3×3 convolutional layer. At last, the spatial-refined matrix is obtained through a sigmoid function, which can be represented as:

$$M_s(F') = \sigma(f^{3 \times 3}(Avg(F'); Max(F'))) \tag{3}$$

Thus the CBAM-refined feature can be obtained by the following formula:

$$F'' = M_s(F') \otimes F' \tag{4}$$

The final change map was learned through metric learning. More specifically, the Euclidean distance between the CBAM-refined feature pairs are first calculated to measure the similarity between the bi-temporal features, then a contrastive loss is employed as the metric to weigh the disparity between the distance map and ground truth. Thereafter, through the optimization of the metric, the distance of the changed area on the ground truth was increased while that of the unchanged area was reduced. In other words, we can obtain distance map with the value difference between the changed area and unchanged area as large as possible through metric learning. Therefore, we could obtain more precise change maps from the distance map by threshold segmentation.

## 3.4. Loss function

There are three sub-models that need to be optimized in our network: the generator, discriminator in the SR module, and change network. As illustrated in the overview, given an LR image $I_{T2}^{LR}$ downsampled from the HR image obtained at T2 $I_{T2}^{HR}$, an SR image $I_{T2}^{SR} = G(I_{T2}^{LR})$ is generated by the generator, which is then transferred to the discriminator and the change detection network successively. Therefore, the optimization of the generator does not precede that of the discriminator and the change network during training.

### 3.4.1. Discriminator

After receiving $I_{T2}^{HR}$ and $I_{T2}^{SR}$ as inputs, the discriminator outputs the probability of the input image being $I_{T2}^{HR}$. To improve the discriminator's ability to accurately distinguish $I_{T2}^{SR}$ from $I_{T2}^{HR}$, the loss function of the discriminator is designed as follows:

$$Loss_D = 1 - D(I_{T2}^{HR}) + D(G(I_{T2}^{LR})) \qquad (5)$$

According to the formula, the discriminator is bound to output a probability close to 1 for an HR image and a probability close to 0 for an SR image after adversarial training.

### 3.4.2. Change detection module

Then, $I_{T2}^{SR}$ is fed to the change detection network together with the HR image obtained at $T1$, $I_{T1}^{HR}$, where the multi-scale features of the bi-temporal inputs are extracted using the Siamese feature extractor. An Euclidean distance map between the feature pairs is calculated based on the feature pairs enhanced by the CBAM block, and based on this map, the final change map is generated using threshold segmentation. Therefore, the objective of the change detection network is to draw the corresponding values of "changed" areas and "unchanged" areas on the distance map $dt$ as much as possible according to the ground truth $gt$. Thus, batch contrastive loss is used to help minimize the distance between "unchanged" areas and maximize the distance between "changed" areas on the distance map, which can be denoted as follows:

$$Loss_{CD} = \sum_{i,j=0}^{M} \frac{1}{2}[(1-gt_{i,j})dt_{i,j}^2 + gt_{i,j}\max(dt_{i,j}-m)^2] \qquad (6)$$

where $M$ denotes the size of $dt$; $dt_{i,j}$ and $gt_{i,j}$ represent the values of the distance map and ground truth map at point *(i, j)*, respectively, where $i, j \in [0, M]$; and m denotes the margin that separates unchanged and changed pixels, which are represented by 0 and 1 on $gt$, respectively.

### 3.4.3. Generator

The generator loss includes the following: image loss, content loss, adversarial loss, and change loss. The image loss measures the alignment of the output SR image $I_{T2}^{SR}$ and the corresponding HR image $I_{T2}^{HR}$ in the pixel-wise space by calculating the mean square errors (MSEs) between them. The image loss can be expressed as:

$$l_{MSE} = \sum_{i,j=0}^{M} (I^{HR}_{T2\ i,j} - G(I^{LR}_{T2})_{i,j})^2 \qquad (7)$$

Since preserving detailed information in an image is difficult as a result of pixel alignment, content loss focuses more on perceptual similarity. Specifically, content loss computes the MSEs between certain feature maps of $I^{SR}_{T2}$ and $I^{HR}_{T2}$ obtained by the pretrained VGG-19 network to yield a more visually realistic SR image. The formula for content loss is:

$$l^{VGG}_{MSE} = \sum_{i,j=0}^{M} (\phi_{VGG}(I^{HR}_{T2})_{i,j} - \phi_{VGG}(G(I^{LR}_{T2}))_{i,j})^2 \qquad (8)$$

While the discriminator is designed to discriminate $I^{SR}_{T2}$ from $I^{HR}_{T2}$, the generator aims to increase the probability of the discriminator's misjudgment through adversarial loss, which can be defined as follows:

$$l_D = 1 - D(G(I^{LR}_{T2})) \qquad (9)$$

To make the SR image pixel-wise and perceptually similar to the original HR image, and to improve at the same time the change detection results, the batch contrastive loss employed in the change detection network is also added to the loss of the generator. In summary, the optimization objective of the generator is:

$$Loss_G = l_{MSE} + \alpha l^{VGG}_{MSE} + \beta l_D + \lambda Loss_{CD} \qquad (10)$$

where $\alpha$, $\beta$ and $\lambda$ are factors that balance different losses.

## 4. Settings

### 4.1. Dataset

In this work, we tested the proposed architecture on two real datasets, which can be summarized as follows.

(1) Building change detection dataset (BCDD): The BCDD [30] provides pairs of 0.2-m images with a size of 32507 × 15354 and a ground truth for the building changes between them. Because the bi-temporal images were selected before and after earthquake occurrence, the areas contain various building changes, including building reconstruction and renewal. For the convenience of model training and testing, we cropped the images into 7434 patches of size 256×256 without overlapping, which were then randomly divided into training, validation, and testing sets in a ratio of 8:1:1. To avoid overfitting, we rotated the images in the training set for data augmentation.

(2) Change detection dataset (CDD): The CDD [31] contains 16000 real season-varying Google Earth image pairs with a size of 256×256, including 10000 training samples, 3000 validation samples, and 3000 testing samples. With a very high spatial resolution of 3 to 100 cm, the CDD not only provides change information of common objects, including buildings, roads, and forests, but also that of many detailed objects, such as cars and tanks, as shown in Fig. 2.

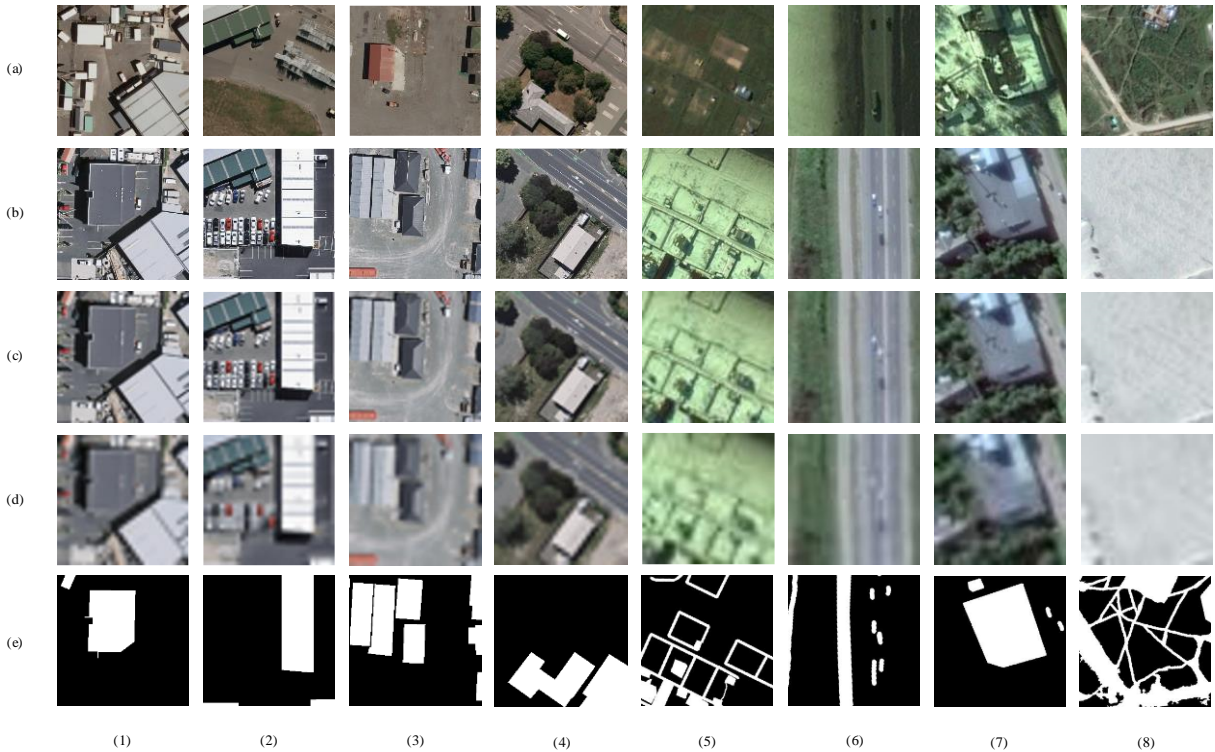

Fig. 2. Examples of the BCDD (1)-(4) and CDD (5)-(8). (a) HR image at time T1; (b) HR image at time T2; (c) 4× LR image at time T2; (d) 8× LR image at time T2; (e) Ground Truth.

## 4.2. Baselines

To validate the proposed SRCDNet, five state-of-the-art (SOTA) change detection methods were introduced into the experiments for comparison. A brief description of each method is provided below.

(1) Fully convolutional–early fusion (FC-EF) [14]: Based on the U-Net architecture, the FC-EF network detects changes by fusing the bi-temporal images as a multispectral input. Skip connections are adopted to transfer features from the encoder to the decoder to recover spatial information at each level.

(2) Fully convolutional–Siamese difference (FC-Siam-diff) [14]: A variant of FC-EF, the FC-Siam-diff network extends the encoder to a Siamese structure to receive bi-temporal inputs and extract their features separately. Features of the same layer in the Siamese encoder are transmitted to the decoder through skip connections after the difference operation.

(3) Fully convolutional–Siamese concatenation (FC-Siam-conc) (Daudt et al. 2018): FC-Siam-conc also adopts

the same Siamese encoder as FC-Siam-diff, and features of the same level in the Siamese encoder are concatenated before being transferred to the decoder, rather than employing the difference operation.

(4) BiDateNet [13]: BiDateNet integrates LSTM blocks into the skip connections of a fully convolutional network with U-Net architecture to learn the temporal dependency between bi-temporal images to help detect changes.

(5) Spatial-temporal attention-based network (STANet) (Chen and Shi 2020): STANet is a metric learning–based change detection network that provides a spatial-temporal attention module to further exploit spatial information and temporal relationships in the features.

## 4.3. Comparative experiments

To fully verify the performance of the proposed method, we designed three groups of comparative experiments according to the scale of the LR images:

(1) No-resolution-difference experiment (X1): This experiment was used to test the performance of the proposed method on images with the same resolution. As mentioned before, the proposed SRCDNet can be flexibly transformed into a common change detection model when there is no resolution difference between the bi-temporal images.

(2) Four-times-resolution-difference experiment (X4): This experiment aims to simulate bi-temporal images with a 4 times resolution difference. Therefore, we downsampled the *T2* images in the BCDD and the CDD four times to obtain LR images. These LR images were then bicubically interpolated to the original image size, to then be used as the input to the comparative methods proposed to deal with images with the same resolution.

(3) Eight-times-resolution-difference experiment (X8): This experiment aims to simulate bi-temporal images with an 8 times resolution difference, which can test the performance of the proposed method under a large spatial difference. The processing of images in the BCDD and CDD is similar to that of experiment X4.

## 4.4. Implementations

We implemented the SRCDNet with PyTorch Libraries. For a total of 100 training epochs, an Adam optimizer with an initial rate of 0.0001 was utilized to facilitate model convergence. A batch size of eight was adopted during training. The $\alpha$, $\beta$ and $\lambda$ in the loss function of the generator are set as 0.006, 0.001, and 0.001, respectively. All of the baselines were run on a GeForce RTX 2080ti graphics card to achieve higher training efficiency.

Four commonly used factors are employed to measure the change detection performance of different methods: precision, recall, F1-score, and IoU. Given that TP, FP, TN, and FN refer to true positives, false positives, true negatives, and false negatives, respectively, precision and recall can be defined as follows:

$$precision = \frac{TP}{TP + FN} \quad (11)$$

$$recall = \frac{TP}{TP + FN} \quad (12)$$

According to the formulas, precision represents the false alarm rate, whereas recall represents the missed alarm rate, both of which entail a trade-off. Thus, to obtain more comprehensive evaluations, the F1 score combines both precision and recall, and can be expressed as follows:

$$F1 = \frac{2 precision \cdot recall}{precision + recall} \quad (12)$$

The IoU refers to the intersection and union rate between the detection result and the ground truth, which can be intuitively represented as:

$$IoU = \frac{TP}{FP + TP + FN} \quad (14)$$

## 5. Experiments and Analysis

### 5.1. *Performance on X1 experiments*

As shown in Table 1, our proposed method outperforms all the baseline methods on the BCDD with the highest recall, F1, and IoU of 90.13%, 87.40%, and 77.63%, respectively, and a very high precision of 84.84%. STANet is ranked second with an F1 and IoU of 84.96% and 73.86%, which are 2.44% and 3.77% lower than those of our method, respectively. BiDateNet obtains the best results among the U-Net-based methods, with an F1 and IoU of 83.55% and 71.75%, respectively, thus proving the feasibility of RNNs for enhancing the time relationship in bi-temporal images. FC-Siam-diff performed the best among the three FCN variants, followed by FC-Siam-conc and FC-EF.

On the CDD, our proposed method and STANet again obtained the best performance, which further proves the superiority of metric learning–based methods compared with traditional classification-based methods. While our proposed method achieves the highest F1 and IoU of 92.94% and 86.81%, STANet obtains a relatively lower F1 and IoU of 91.44% and 84.23%, respectively, which shows the effectiveness of the stacked attention module. The third-ranked BiDataNet had the highest precision of 95.98%. In contrast to the results on the BCDD, FC-Siam-conc performed better than FC-Siam-diff, with increases in F1 and IoU of 2.97% and 4.44%, respectively. This may be attributed to the fact that because the CDD contains various fine-grained objects, much useful information is over-filtered through the difference operation of bi-temporal features in skip connections, whereas concatenation can better save such information in features.

The above results are further verified by visualization comparisons, as shown in Fig. 3, where the proposed method obtains the best visualization performance. All methods can accurately extract newly built buildings in the BCDD. In addition, the FC-EF and FC-Siam-diff entail many missed alarms, as shown in the first row of Fig. 3; this corresponds to the low recall of the two methods, shown in Table 1. Compared with other methods, our proposed method is able to capture more precise building footprints, which is of great significance for practical applications. It is also important to note that our method is the only one to detect a building reduction, as shown in the second row of Fig. 3.

As for the CDD, only changes corresponding to large areas can be extracted by the three FCN variants, explaining their high precision and low recall, as shown in Table 1. Benefitting from the integration of LSTM blocks, BiDateNet is better able to extract small changes compared with the other U-Net-based methods. The metric learning–based methods are good at capturing small changes, including cars and roads, as shown in Fig. 3. The STANet can extract the greatest number of small changes; however, the results show a spillover effect, which leads to its high recall but low precision, as shown in Table 1. In terms of visualization results, our proposed method can not only extract small changes precisely but also maintain their boundaries and shapes in a better way.

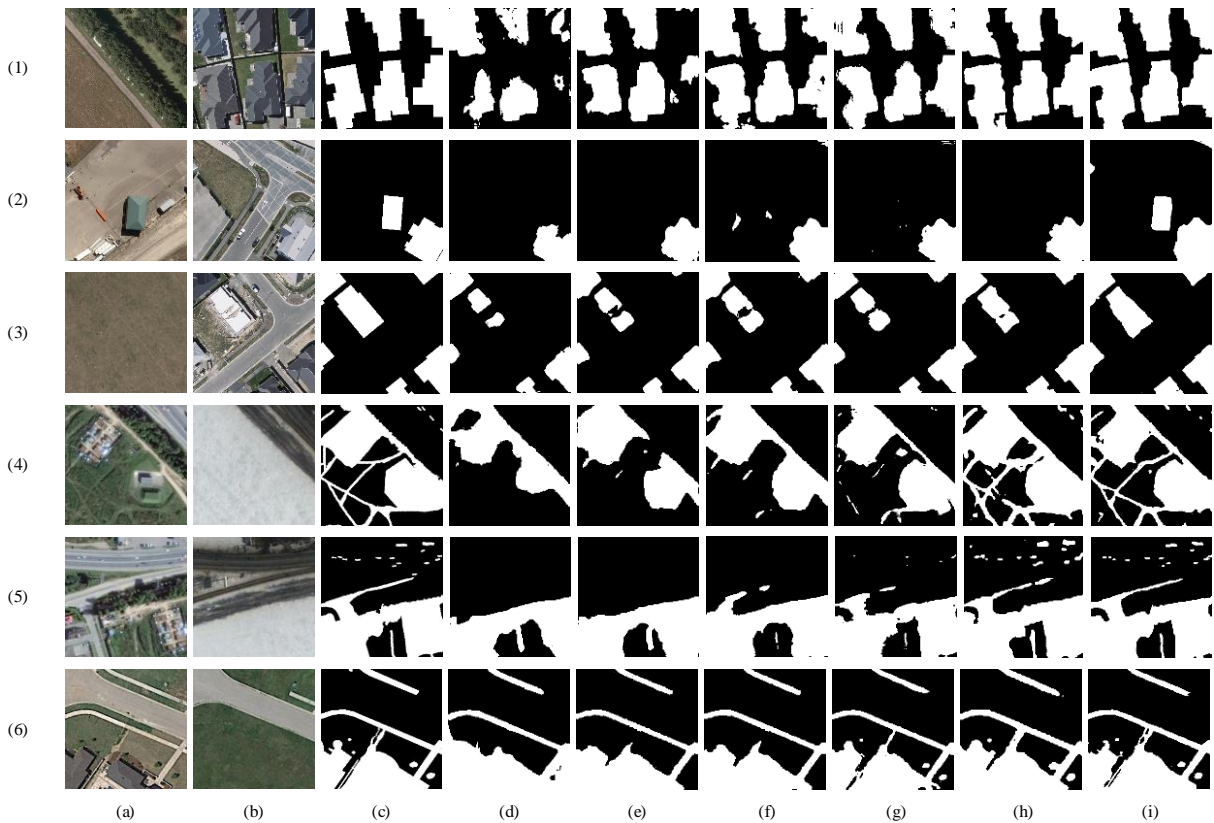

**Fig. 3.** Visualization comparisons on X1 experiments (BCDD: (1)-(3); CDD: (4)-(6)). (a) HR image at T1; (b) HR image at T2; (c) ground truth; (d) FC-EF; (e) FC-Siam-diff; (f) FC-Siam-conc; (g) BiDateNet; (h) STANet; (i) proposed method.

## 5.2. Performance on X4 experiments

As shown in Table 1, when there is a 4× resolution difference between the bi-temporal images, both the F1 and IoU of all methods decline slightly on the BCDD dataset compared with those in the X1 experiments. More specifically, SRCDNet obtains the highest F1 and IoU of 85.66% and 74.91% on the BCDD, which are 1.74% and 2.72% lower than those obtained in the X1 experiments. Although STANet outperformed BiDateNet in the X1 experiment, they achieved similar results in the X4 experiment, with F1 scores of 81.96% and 81.97%, respectively, suggesting that STANet is more affected by the resolution difference. Among the three FCN variants, FC-Siam-diff achieved the best performance, followed by FC-Siam-conc and finally FC-EF, which is consistent with the results of the X1 experiments. The accuracies of all baselines decreased more significantly on the CDD. FC-Siam-conc had the worst detection results, with an F1 of 73.67%, which is 12.23% lower than that of the X1 experiment. The F1 of FC-Siam-diff was slightly higher, at 76.15%. Notably, FC-EF obtained the highest F1 of 76.58% among the three FCN variants, only 2.07% lower than that of the X1 experiment. BiDateNet and STANet have very close F1 values of 86.28% and 86.49%, which are 3.73% and 4.95% lower than those of the X1 experiment, respectively. In this case, SRCDNET still obtained a very high F1 and IoU of 90.02% and 81.86%, with only a 2.92% reduction in F1.

The visualization comparisons of these methods are shown in Fig. 4. Most of the building changes in the BCDD can be detected relatively completely from the 4× bicubic images, which indicates that owing to the large size of the buildings, the 4× resolution difference can be mitigated to some extent by bicubic interpolation. This also explains that the accuracy of the X4 experiment only exhibits a small decline compared with that of the X1 experiment. However, the boundaries of the building changes obtained based on the comparative methods are not sufficiently regular and smooth, in contrast with those obtained via SRCDNet. This can be attributed to the fact that the SR image generated by the SRCDNet can better restore the boundary information of buildings compared with the bicubic image. In addition, while some bare lands adjacent to the building are easily extracted as part of the building changes, the SRCDNet can effectively reduce these pseudo-changes, as shown in Fig. 4-(3).

Compared to the BCDD, the CDD has a higher spatial resolution and thus contains more detailed changes, which are difficult to reconstruct in 4× bicubic images; thus, the accuracy of the comparative methods declined significantly compared with that obtained in the X1 experiment. While some detailed changes -such as alleys in Fig. 4-(6) and cars in Fig. 4-(7) are rarely seen in the results of comparative methods, the SRCDNet is the only method to detect such changes, which further proves the validity of the SR module with regard to recovering spatial and semantic information. In addition to extracting smaller changes, SRCDNet also exhibits better performance in extracting large-area changes such as land changes and building changes. Thus, the quantitative and visualization comparisons fully demonstrated the effectiveness of SRCDNet on 4× LR images of both datasets.

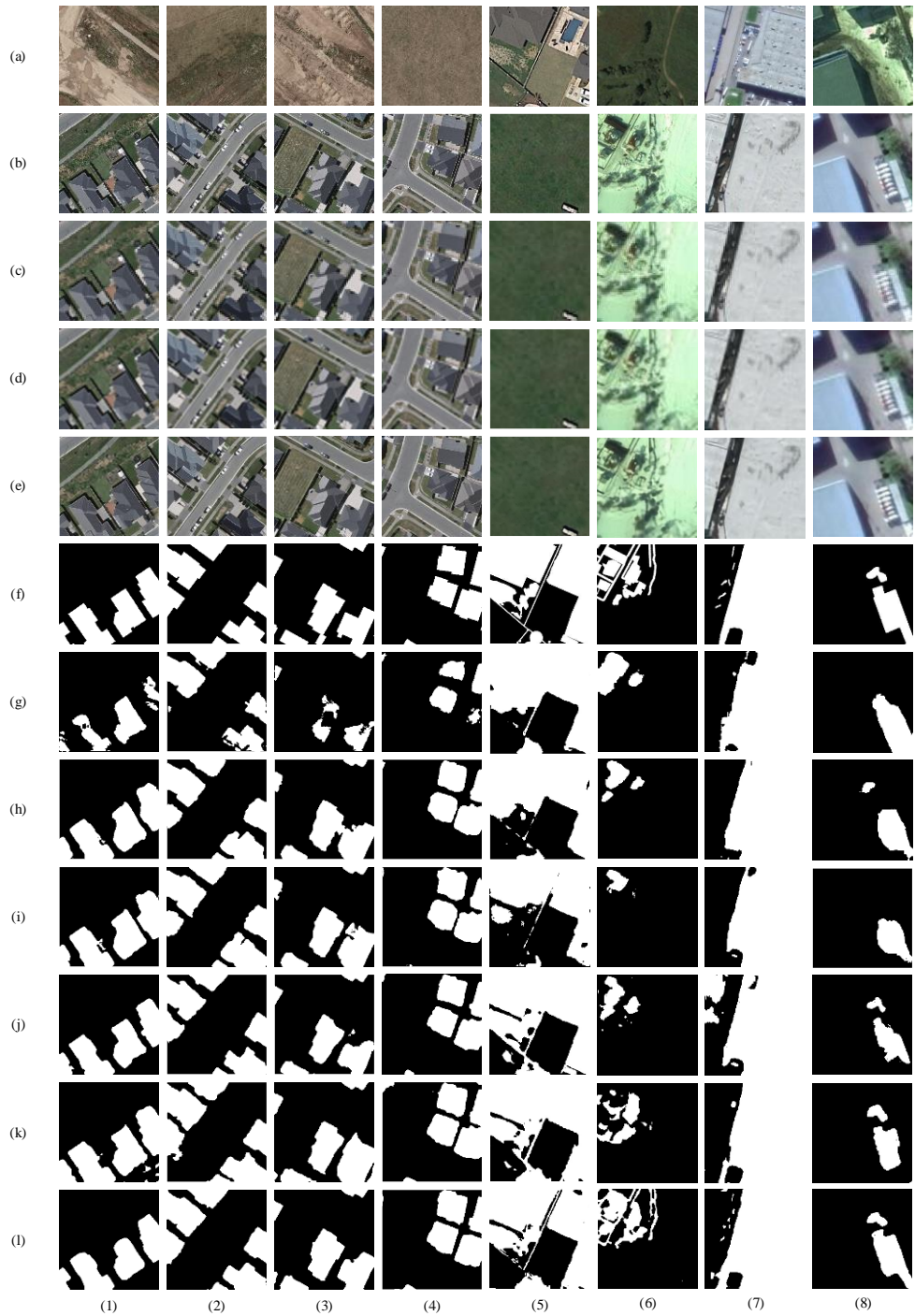

**Fig. 4.** Visualization comparisons on X4 experiments (BCDD: (1)-(4); CDD: (5)-(8)). (a) HR image at time T1; (b) HR image at time T2; (c) 4× LR image at time T2; (d) 4× Bicubic image at time T2; (e) 4× SR image at time T2; (f) ground truth. (g) FC-EF; (h) FC-Siam-diff; (i) FC-Siam-conc; (j) BiDateNet; (k) STANet; (l) SRCDNet.

## 5.3. Performance on X8 experiments

The X8 experiment compares the performance of all the baselines on 8× LR images, shown in Table 1.

On the BCDD, the accuracies of all methods were further reduced compared with the X4 experiment. FC-Siam-conc had the highest accuracy reduction, with the lowest F1 of 69.01%, followed by BiDateNet with an F1 of 70.01%. The F1 of FC-EF decreased from 76.69% to 73.49%, and it was least affected by the resolution difference. FC-Siam-diff achieved the highest F1 accuracy among the U-Net-based models, with an F1 of 77.21%. STANet obtained an F1 of 77.31% and an IoU of 69.05%. Despite the large resolution differences, SRCDNet outperformed all the comparative methods and obtained the highest F1 of 81.69% and IoU of 69.05%.

It is difficult to recover many small objects from LR images, and thus, the accuracies obtained on the CDD are lower than those obtained on the BCDD. More specifically, SRCDNet achieved the highest detection rate, with an F1 of 83.32% and an IoU of 71.40%, which are 6.70% and 10.46% lower than those obtained in the X4 experiments, respectively. Additionally, the F1 of BiDateNet decreased from 86.28% to 78.29%, whereas that of STANet decreased from 86.49% to 77.29%. Owing to the poor ability of the three FCN variants to extract small changes, the accuracies of these three methods were not greatly reduced in the X8 experiment.

The visualization results of the X8 experiments are shown in Fig. 5. Taking into account the results on the BCDD in Fig. 5-(1)-(4), it is possible to appreciate that , it is difficult to obtain regular building boundaries based on the bicubic images, as a result of the large resolution difference. Moreover, there are also a large number of missed alarms; that is, many houses with small volumes are missed. However, SRCDNet can better solve the aforementioned problems and generate more precise change results. It can be seen that despite the 8× resolution difference, the SR image can still recover the information of the building well, thanks to sufficient prior knowledge.

Numerous missed alarms also occur on the CDD, which is consistent with the low recall rates of all methods, as shown in Table 1. This is because it is difficult to fully recover the information in the initial HR images, which had been greatly reduced after 8× downsampling, by using bicubic interpolation. Therefore, not only small changes, including those of alleys and cars, but also some building changes are missed or incompletely detected. Notably, compared with the bicubic image, the SR image output by SRCDNet better regains information from the 8× LR image, which significantly helps in subsequent change detection, leading to more complete and accurate change results.

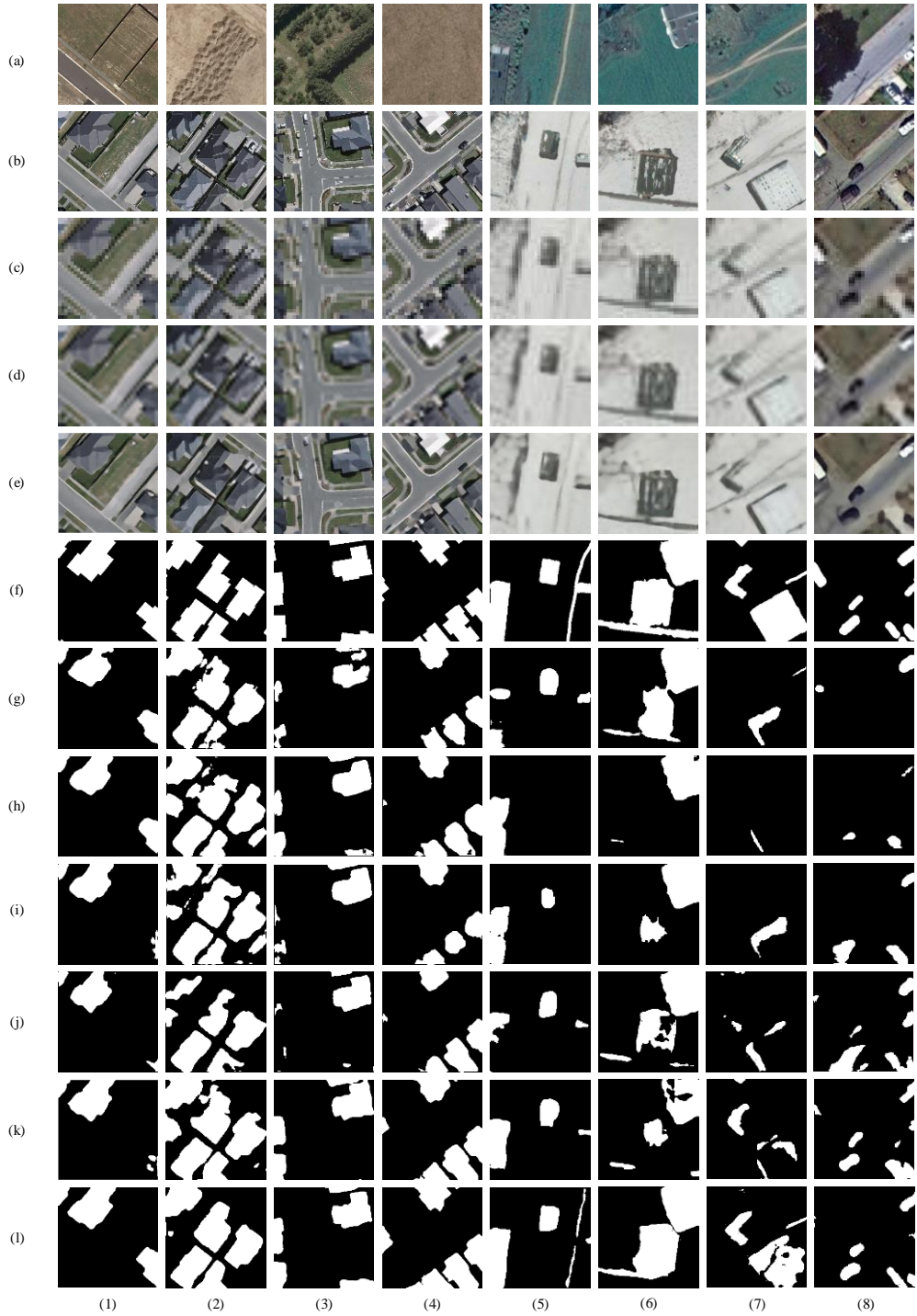

Fig. 5. Visualization comparisons on X8 experiments (BCDD: (1)-(4); CDD: (5)-(8)). (a) HR image at time T1; (b) HR image at time T2; (c) 8× LR image at time T2; (d) 8× Bicubic image at time T2; (e) 8× SR image at time T2; (f) ground truth. (g) FC-EF; (h) FC-Siam-diff; (i) FC-Siam-conc; (j) BiDateNet; (k) STANet; (l) SRCDNet.

# 6. Discussion

To verify the effectiveness of the stacked attention module (SAM) and super-resolution module (SRM), we conducted an ablation study on SRCDNet with two resolution differences (X4 and X8) on both datasets. The SRCDNet without SAM and SRM was set as the base model. Then, the SAM was added to the base model as the second baseline, and the SRM was added to the base model as the third baseline. The resultant image of the bicubic interpolation of the LR image is used as the input because the first two baselines do not contain an SRM. The quantitative results of the ablation study on SRCDNet are summarized in Table 2.

## 6.1. Ablation study with 4× resolution difference

Table 2 Ablation Study on SRCDNet.

| Dataset | Baseline | X4 | | | | X8 | | | |
|---|---|---|---|---|---|---|---|---|---|
| | | Pre(%) | Rec(%) | F1(%) | IoU(%) | Pre(%) | Rec(%) | F1(%) | IoU(%) |
| BCDD | Base | 67.20 | **91.63** | 77.53 | 63.31 | 62.63 | 85.40 | 72.26 | 56.57 |
| | Base+SAM | 80.43 | 88.06 | 84.07 | 72.52 | 70.85 | 84.84 | 77.22 | 62.89 |
| | Base+SRM | 72.72 | 88.42 | 79.81 | 66.40 | 65.73 | **87.77** | 75.17 | 60.22 |
| | SRCDNet | **84.44** | 86.90 | **85.66** | **74.91** | **81.61** | 81.78 | **81.69** | **69.05** |
| CDD | Base | 90.03 | 83.35 | 86.56 | 76.31 | 87.08 | 70.62 | 77.99 | 63.92 |
| | Base+SAM | **93.31** | 84.66 | 88.77 | 79.82 | 91.41 | 70.97 | 79.91 | 66.54 |
| | Base+SRM | 93.03 | 82.65 | 87.53 | 77.83 | 88.85 | 73.74 | 80.59 | 67.49 |
| | SRCDNet | 92.07 | **88.07** | **90.02** | **81.86** | **91.95** | **76.03** | **83.24** | **71.29** |

As shown in Table 2, the base model without any additions performs the worst on both datasets, with the lowest F1 values of 77.53% on the BCDD and 86.56% on the CDD. With the integration of the SAM, the F1 on the BCDD was significantly increased to 84.07%, and that on the CDD also increased to 88.77%. Notably, the precision rates of the second baseline are greatly improved compared with those of the base model, which shows that the SAM can help extract changes more accurately. Compared with the base model, the third baseline that included the SRM also obtained better detection results, with an F1 of 79.81% on the BCDD and an F1 of 87.53% on the CDD. Therefore, SRCDNet, which integrates both the SAM and SRM into the base model, outperforms all baselines, with the highest F1 values of 85.66% on the BCDD and 90.02% on the CDD.

As shown in Fig. 6, the performance of the base model is poor on both datasets. More specifically, the building changes in the BCDD have an obvious spillover effect, whereas the change results corresponding to the CDD are substantially missed. The second baseline can solve the above problems and greatly improve the detection accuracy owing to the SAM. With the inclusion of the SRM to generate finer-grained SR images for change detection, the results of the third baseline have more precise change boundaries; however, this is not accurate enough because of the poor detection performance of the base model. Therefore, by combining the advantages of the SAM and SRM, SRCDNet can obtain the most precise change maps among all the baselines on the two datasets, based on the output SR images.

### 6.2. Ablation study with 8× resolution difference

The ablation study on 8× LR images on the BCDD showed the same tendency as that on 4× LR images. While the base model could only obtain an F1 of 72.26%, the SAM-integrated model was able to boost the F1 to 77.22%, and the SRM-integrated model could improve it to 75.17%. SRCDNet far surpasses the above baselines, with an F1 of 81.69% and an IoU of 69.05%, which fully demonstrates the feasibility of combining SAM and SRM.

The performance of the base model on the CDD is the worst among all baselines, with an F1 of 77.99%. Whereas SAM integration was more effective than SRM integration in the previous experiments, the opposite effect is observed on the CDD for 8× LR images. More specifically, the F1 of the second baseline with the SAM increased to 79.91%, whereas that of the third baseline increased to 80.59% owing to the addition of the SRM. This may be because the SRM can enhance the extraction of small changes in the images in the CDD by effectively recovering image information. Nevertheless, SRCDNet has the highest F1 of 83.24%, which is much higher than that of the other baselines.

According to Fig. 7, compared with the base and base+SAM models based on bicubic images, the base+SRM model and SRCDNet can better capture the building boundaries on the BCDD owing to the integration of the SRM. Moreover, they can significantly alleviate missed alarms owing to SR images with more detailed information, as shown in Fig. 7-(2). Because the information of small objects in the CDD is difficult to recover from bicubic images, the change results obtained on the CDD by the base and base+SAM models also entail several missed alarms. Although it is difficult to restore the 8× LR images, with the help of the SRM, the base+SRM and SRCDNet models yield more comprehensive change results.

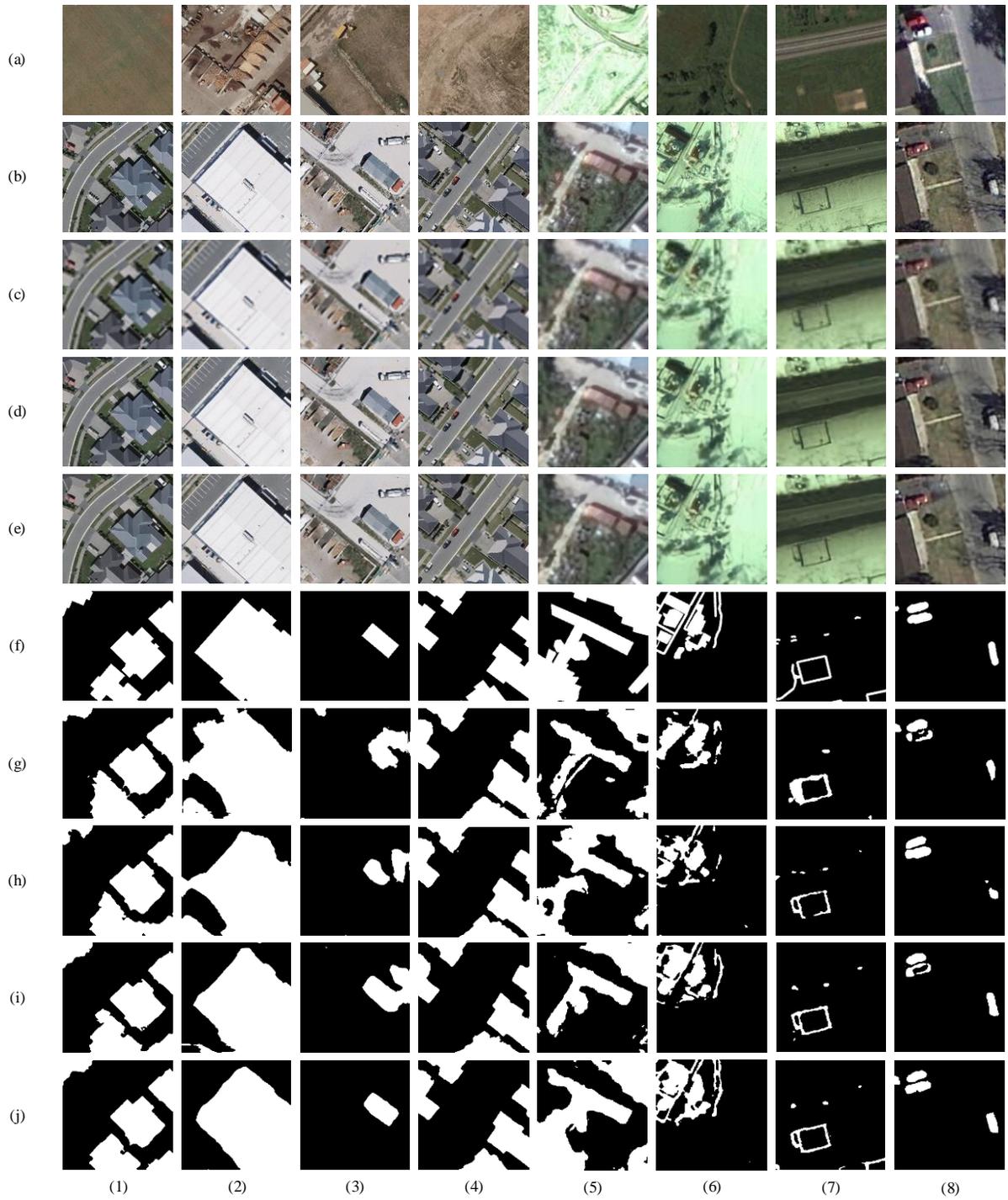

**Fig. 6.** Ablation study on X4 images (BCDD: (1)-(4); CDD: (5)-(8)). (a) HR image at time T1; (b) HR image at time T2; (c) 4× Bicubic image at time T2; (d) 4× SR image at time T2 by Base+SRM; (e) 4× SR image at time T2 by SRCDNet; (f) ground truth; (g) Base; (h) Base+SAM; (i) Base+SRM; (j) SRCDNet.

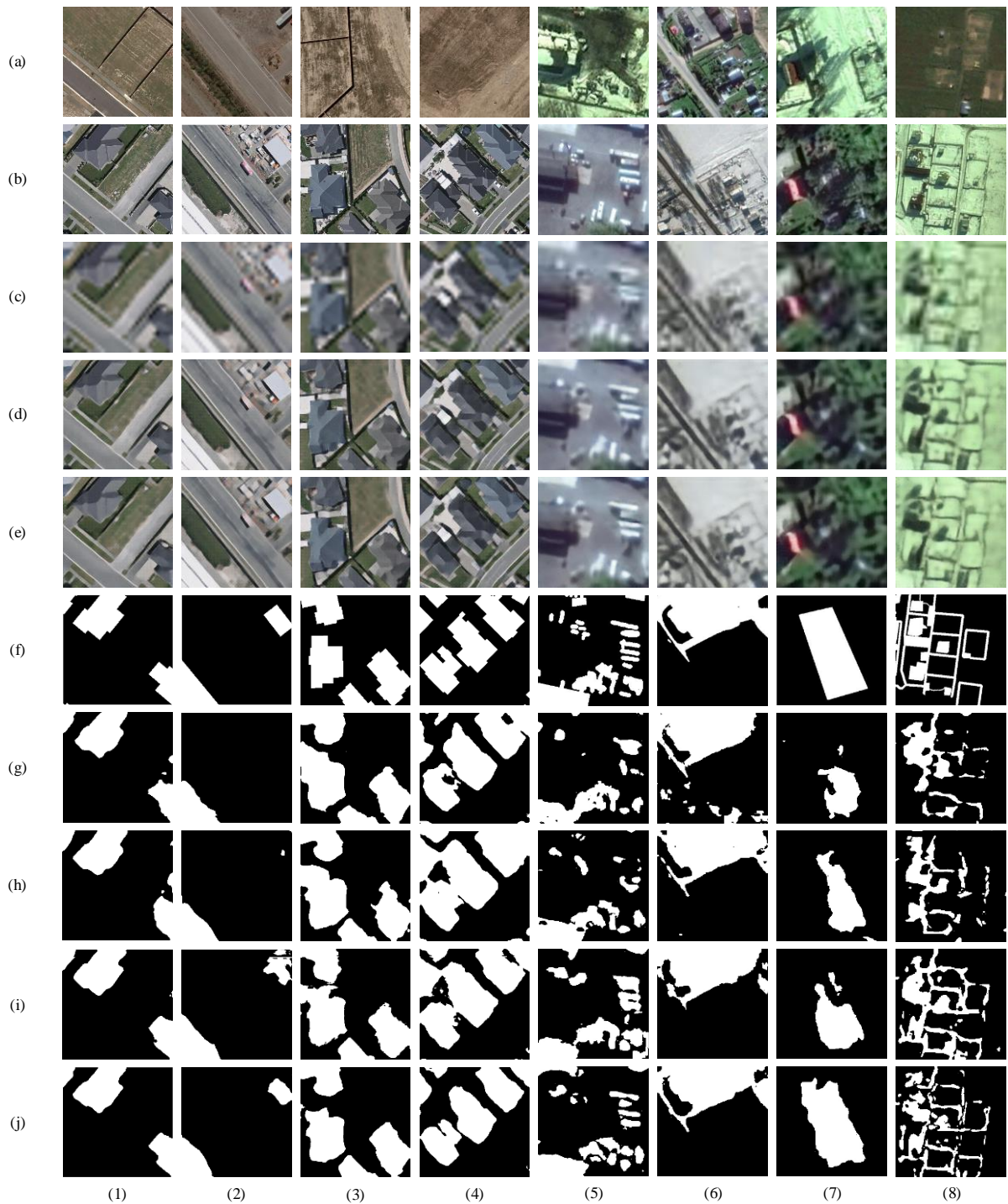

Fig. 7. Ablation study on X8 images (BCDD: (1)-(4); CDD: (5)-(8)). (a) HR image at time T1; (b) HR image at time T2; (c) 8× Bicubic image at time T2; (d) 8× SR image at time T2 by Base+SRM; (e) 8× SR image at time T2 by SRCDNet; (f) ground truth; (g) Base; (h) Base+SAM; (i) Base+SRM; (j) SRCDNet.

## 6.3. Comparisons on different restored images

According to the ablation study presented above, two different methods, bicubic interpolation and SR, can be utilized to restore detailed information from LR images. While both the Base+SRM model and SRCDNet have adopted the SRM to generate the SR image from the LR image, there are three sets of restored images for change detection for each group of experiments. Hence, we employ two common indexes, namely, peak signal-to-noise ratio (PSNR) and structural similarity (SSIM), to compare the effects of different restored images quantitatively and further understand the effect of the different modules in SRCDNet.

As can be seen from Table 3, the bicubic image obtains the lowest PSNR and SSIM in each comparative experiment. Further, the indexes of the SR image obtained using the base+SRM model are improved to a certain extent, which illustrates that the SRM can produce restored images with higher quality compared with bicubic interpolation. The SR image obtained using SRCDNet achieves the highest PSNR and SSIM in all experiments, with slightly higher metrics than those of the SR image obtained using the base+SRM model. This further proves that the combination of SAM and SRM has a positive effect on change detection, as mentioned before.

Table 3 Comparisons on different restored images.

| Baselines | Index | BCDD | | CDD | |
|---|---|---|---|---|---|
| | | X4 | X8 | X4 | X8 |
| Bicubic Image | PSNR | 21.87, | 20.25, | 28.95 | 25.23 |
| | SSIM | 0.5051 | 0.3774 | 0.7571 | 0.6408 |
| SR Image | PSNR | 23.04, | 21.03, | 29.63, | 25.83, |
| (Base+SRM) | SSIM | 0.5920 | 0.4322 | 0.7811 | 0.6580 |
| SR Image | PSNR | **23.09,** | **21.04,** | **29.88,** | **25.89,** |
| (SRCDNet) | SSIM | **0.5929** | **0.4335** | **0.7818** | **0.6582** |

Another noteworthy phenomenon is that in the X4 experiments, the integration of the SRM leads to a greater increase in the PSNR and SSIM. Hence, more image information can be recovered through the SRM compared with the case of the X8 experiment. This shows that the greater the resolution difference, the greater the loss of image information, which causes substantial difficulties in image restoration.

# 7. Conclusion

We propose an end-to-end super-resolution-based change detection network (SRCDNet) for bi-temporal images at different scales. With the aim of overcoming the resolution difference between bi-temporal images, a SR module consisting of a generator and a discriminator is adopted to restore LR images to the size of HR images, which has been proven to be effective in generating realistic SR images from LR images. A Siamese feature extractor extracts multi-scale features from two input images, an SR image and a HR image corresponding to a different timestamp; subsequently, a stacked attention module is applied to capture more useful channel-wise and spatial information. SRCDNet also employs deep metric learning to learn the final change map. Comparative experiments were conducted on the BCDD and CDD to test the effectiveness of SRCDNet. Compared with the SOTA baselines, the proposed SRCDNet obtains the best results for images with the same resolution. Furthermore, it outperforms other comparative methods in change detection under different resolutions, providing a more general solution for multi-scene change detection.

**Acknowledgments**

This work was supported by Guangdong Natural Science Foundation (2019A1515011057), National Natural Science Foundation of China (61976234) and Open research fund of National Key Laboratory of surveying, mapping and remote sensing information engineering, Wuhan University, Guangzhou Applied Basic Research Project, by Centre for Integrated Remote Sensing and Forecasting for Arctic Operations (CIRFA) and the Research Council of Norway (RCN Grant no. 237906).